\documentclass[11pt]{article}

\usepackage{graphicx}
\def\real{{{\rm I}\!{\rm R}}}

\title{Symmetry of Planar Four-Body \\
Convex Central Configurations }

\author{ Alain Albouy$^{1}$,\thanks{E-mail: albouy@imcce.fr
 }\qquad
 Yanning
Fu$^{2}$,\thanks{Supported by NSFC (grant numbers 10473025 and
10233020). E-mail: fyn@pmo.ac.cn }\qquad Shanzhong Sun$^{1,3}$\thanks{Partially
supported by NSFC (grant numbers 10401025 and 10571123) and by NSFB-FBEC(grant number
KZ200610028015). E-mail: sunsz@mail.cnu.edu.cn}\\ \\
$^{1}$ CNRS-UMR 8028, Observatoire de Paris\\
  77, avenue Denfert-Rochereau\\
  75014 Paris, France\\
$^{2}$ Purple Mountain Observatory\\
2 West Beijing Road\\
Nanjing 210008, P. R. China\\
$^{3}$ Department of Mathematics, Capital Normal University,\\
Beijing, 100037, P. R. China}

\date{}

\begin{document}

\maketitle

\begin{abstract}
We study the relationship between the masses and the
geometric properties of central configurations. We
prove that in the planar four-body problem, a
convex central configuration is symmetric with respect to one
diagonal if and only if the masses of the two particles on the other
diagonal are equal. If these two masses are unequal, then the less massive one is closer to the former diagonal.
Finally, we extend these results to the case of non-planar central configurations of five particles.

\end{abstract}

{\bf 1. Introduction.} Let $q_1,\dots, q_n$ represent the positions in a
Euclidean space
$E$ of $n$ particles with respective positive masses
$m_1,\dots,m_n$. Let $$\gamma_i=\sum_{k\neq i}m_k S_{ik}
(q_i-q_k),\qquad S_{ik}=S_{ki}=\|q_i-q_k\|^{2a},\qquad
a=-3/2.\eqno(1)$$ Newton's equations
$$\frac{d^2q_i}{dt^2}+\gamma_i=0\eqno(2)$$
define the classical $n$-body motions of the system of particles. In
the Newtonian $n$-body problem, the simplest possible motions are
such that the configuration is constant up to rotation and scaling,
and that each body describes a Keplerian orbit. Only some special
configurations of particles are allowed in such motions. Wintner
called them ``central configurations''. Famous authors as
Euler \cite{Eu}, Lagrange \cite{Lag}, Laplace \cite{Laplace},
Liouville \cite{Lio}, Maxwell \cite{Maxwell} initiated the study
of central configurations, while Chazy \cite{Cha},
Wintner \cite{Wintner}, Smale \cite{Smale} and
Atiyah\&Sutcliffe \cite{Atiyah} called attention to hard unsolved
questions. Chazy, Wintner and Smale conjectured that the number of central configurations
of $n$ particles with given masses is finite. This was proved recently in the case $n=4$ by Hampton and Moeckel \cite{HaM}.
This reference also provides an excellent review, which describes how central configurations appear in several applications. Let
us mention that central configurations are also needed in the application of Ziglin method or the computation of Kowalevski
exponents, in order to get statements of non-existence of additional first integrals in the $n$-body problem
(see e.g.\ Tsygvintsev\cite{Tsygv}).

In terms of $$M=m_1+\cdots+m_n,\qquad
q_G=\frac{1}{M}(m_1q_1+\cdots+m_nq_n),\eqno(3)$$ a configuration
$(q_1,\dots,q_n)$ is called a central configuration if and only if
there exists a $\lambda\in\real$ such that
$$\gamma_i=\lambda(q_i-q_G).\eqno(4)$$
The results in this note are not restricted to the Newtonian case
$a=-3/2$. They are true for any $a<0$. In particular, they apply to
relative equilibria of Helmholtz vortices in the plane with positive
vorticities $m_1,\dots,m_n$ (see e.g.\ O'Neil \cite{ONeil}), for which
$a=-1$. While vorticities may be negative, we always assume $m_i>0$ for all $i$, $1\leq i\leq n$. Note that it implies
$\lambda>0$. For results with some negative masses see for example Celli \cite{Celli}. The
main result we prove in this note is:

{\bf Theorem 1.} Let four particles $(q_1,q_2,q_3,q_4)$ form a
two-dimensional central configuration, which is a convex
quadrilateral having $[q_1,q_2]$ and $[q_3,q_4]$ as diagonals (see
Fig.\ 1). This configuration is symmetric with respect to the axis
$[q_3,q_4]$ if and only if $m_1=m_2$. It is  symmetric with respect
to the axis $[q_1,q_2]$ if and only if $m_3=m_4$. Also $m_1<m_2$ if
and only if $|\Delta_{134}|<|\Delta_{234}|$, where $\Delta_{ijk}$ is
the oriented area of the triangle $[q_i,q_j,q_k]$. Similarly,
$m_3<m_4$ if and only if $|\Delta_{123}|<|\Delta_{124}|$.

{\sl Remark 1.} In Theorem 1, each geometric relation between areas has equivalents in terms of distances.
For example, $|\Delta_{134}| < |\Delta_{234}|$ is obviously equivalent to that $q_1$ is closer to the line $[q_3,q_4]$
than $q_2$. By Lemma 1 and $(8)$, it is also equivalent to $\|q_1-q_3\|<\|q_2-q_3\|$ or to $\|q_1-q_4\|<\|q_2-q_4\|$.

{\sl Remark 2.} Our hypotheses are not far from optimal. We assume $a<0$ in $(1)$ being aware of counter-examples if $a>0$,
and of course if $a=0$. If the equal masses are $m_1$ and $m_3$ there is no symmetry except if also $m_2=m_4$ (but no proof is
known of the symmetry in this latter case). The convexity of the configuration is also a necessary hypothesis: there are
asymmetric central configurations where a particle $q_1$ is inside the triangle formed by $q_2$, $q_3$ and $q_4$, and where
$m_3=m_4$. This can be seen immediately by perturbing the four equal mass
solution.

\begin{figure*} %4
\centerline{\includegraphics [width=110mm] {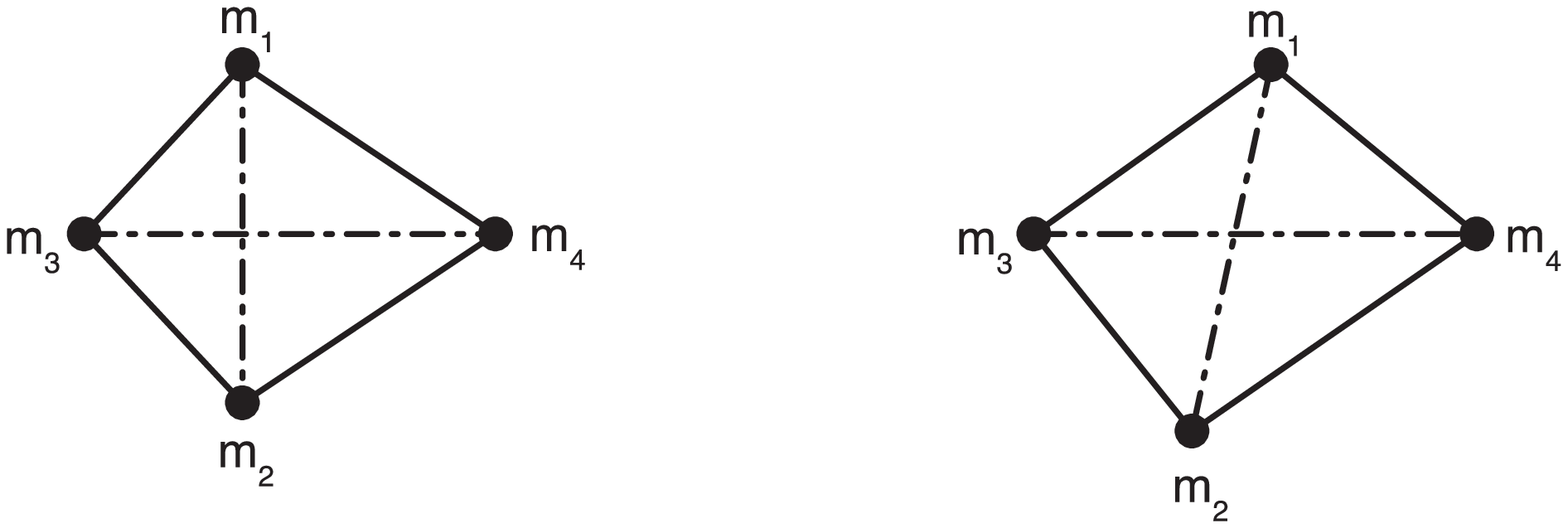}}
\caption{Planar 4-body convex configurations}
\centerline{Left: with an axis of symmetry. Right: asymmetric.}
\end{figure*}

For any choice of four positive masses there is always a convex central configuration with the particles ordered as in Figure 1,
according to \cite{MacMillan-Bartky, Xia}. Theorem 1 associated to Leandro's \cite{Le}
shows the uniqueness (up to trivial transformations) in the case where two particles with equal masses are on a diagonal. We
conjecture that the uniqueness of the convex central configuration is true for any choice of four positive masses (see also
\cite{AlbouyFu} which finishes with a selection of open questions).

Given the presence of some equal masses, it is natural to attack this conjecture by proving
some symmetry of the central configuration at first. In this
direction, the first named author took the first step \cite{Albouy1,Albouy2}. He
studied the equal mass case and gave a complete understanding on the
geometric properties and the enumeration of the central configurations
convex or not. Then Long and the third named author \cite{LS}
studied the convex central configurations with $m_1=m_2$ and $m_3=m_4$,
and they proved symmetry and uniqueness under some constraints which are
dropped out by Perez-Chavela and Santoprete. In
\cite{PeS}, they also generalise further and get symmetry on convex central
configurations with $m_1=m_2$ only, and the uniqueness follows
Leandro \cite{Le}. Unfortunately, in this situation, Perez-Chavela and Santoprete
also need that $m_1=m_2$ are not the smallest masses. Our Theorem 1 gives the
expected symmetry assuming simply $0<m_1=m_2$, $0<m_3$, $0<m_4$, and
choosing any $a<0$ in Equation $(1)$. The uniqueness in the Newtonian
case $a=-3/2$ follows Leandro \cite{Le}.

{\bf 2. Dziobek's equations.} The main known results on the planar central
configurations of four bodies were obtained or simplified using
Dziobek's coordinates \cite{Dziobek}. In particular, Meyer and Schmidt
\cite {MeS} remarked their effectiveness, and extended their use to the
non-planar central configurations of five bodies.

Given a $(n-2)$-dimensional configuration $(q_1,\dots,q_n)$ of $n$ particles, we consider the linear equations
$$\Delta_1+\cdots+\Delta_n=0,\qquad \Delta_1q_1+\cdots+\Delta_nq_n=0.\eqno(5)$$
A non-zero $(\Delta_1,\dots,\Delta_n)\in\real^n$ satisfying $(5)$
will be called a system of {\sl homogeneous barycentric coordinates}
of the configuration. Such a system is not unique but defined up to
a non-zero real factor. There is a $\kappa\in\real$ such that
$\pm\kappa\Delta_i$  is the $(n-2)$-dimensional volume of the
simplex $(q_1,\dots,q_{i-1},q_{i+1},\dots,q_n)$. This remark
is useful for the geometrical intuition, but at the technical level
it only introduces complications. We rewrite the equations for
central configuration $(4)$ as $$0=\sum_{k\neq i}m_k\check
S_{ik}(q_k-q_i),\qquad\hbox{with}\qquad \check
S_{ik}=S_{ik}-\frac{\lambda}{M}.\eqno(6)$$ and compare them with the
second equation of $(5)$ written as
$$\sum_{k\neq i}\Delta_k(q_i-q_k)=0.\eqno(7)$$
By the uniqueness, up to a factor, of $(\Delta_1,\dots,\Delta_n)$,
there exists a real number $\theta_i$ such that
$\theta_i\Delta_k=m_k\check S_{ik}$. We can write $\check
S_{ik}=\check S_{ki}=\theta_i\Delta_k/m_k=\theta_k\Delta_i/m_i$ for
any $(i,k)$, $1\leq i<k\leq n$. Thus $(\theta_1,\dots,\theta_n)$ is
proportional to $(\Delta_1/m_1,\dots,\Delta_n/m_n)$. Calling $\mu$
the proportionality factor, we obtain Dziobek's equations
$$S_{ij}-\frac{\lambda}{M}=\mu\frac{\Delta_i\Delta_j}{m_im_j}.\eqno (8)$$

{\bf 3. Routh versus Dziobek.}  Here we only consider the case
$n=4$. Equation $(8)$ was proved by Dziobek by
considering Cayley's determinant $H$, differentiating $H$ on $H=0$
and characterising the central configurations as critical points of
the Newtonian potential restricted to some submanifold of $H=0$. The expression of the derivatives of $H$ is quite nice, but
the computation giving it is not so easy. Dziobek and many authors after him chose to skip it. An elegant presentation is
given by Moeckel \cite{Moeckel}.

The ``vectorial" deduction of $(8)$ we just gave is simpler than
Dziobek's approach. It was also published in \cite{Moeckel}. Previously many authors presented similar vectorial
computations and deduced relations which are easy consequences of
Dziobek's equations. A formula as
$$m_3\Delta_{123}(S_{13}-S_{23})+m_4\Delta_{124}(S_{14}-S_{24})=0\eqno(9)$$
appeared in Routh \cite{Routh1} and Krediet \cite{Krediet} before Dziobek's
work, in Laura  \cite{Laura} and Andoyer \cite{And} just after. The
symbol $\Delta_{ijk}$ represents the oriented area of the triangle
$(i,j,k)$. Setting $\Delta_4=\Delta_{123}$ and
$\Delta_3=-\Delta_{124}$, we recognise $(9)$ as an easy consequence of $(8)$.
But $(8)$ is not an obvious corollary of $(9)$, and none of these four authors wrote it.

In the last edition \cite{Routh2} of his treatise, 28 years after \cite{Routh1}, Routh considered again the central
configurations. Apparently he wanted to claim priority against someone, and it is very likely that this someone is Dziobek.
Routh uses Dziobek's notation $\Delta$, which suggests strongly that he read Dziobek's paper. Routh's
priority claim concerns $(9)$ and another important formula. We reproduce the text in \cite{Routh1},
which is not easy to find in the libraries.

{\small
 ``Ex.\ 2. If four particles be placed at the corners of a
quadrilateral whose sides taken in order are $a$, $b$, $c$, $d$ and
diagonals $\rho$, $\rho'$, then the particles could not move under
their mutual attractions so as to remain always at the corners of a
similar quadrilateral unless
$$(\rho^n\rho'^n-b^nd^n)(c^n+a^n)+(a^nc^n-\rho^n\rho'^n)(b^n+d^n)+(b^nd^n-a^nc^n)(\rho^n+\rho'^n)=0,$$
where the law of attraction is the inverse $(n-1)^{\rm th}$ power of the distance.

Show also that the mass at the intersection of $b$, $c$ divided by
the mass at intersection of $c$, $d$, is equal to the product of the
area formed by $a$, $\rho'$, $d$ divided by the area formed by $a$,
$b$, $\rho$ and the difference $\frac{1}{\rho'^n}-\frac{1}{d^n}$
divided by the difference $\frac{1}{\rho^n}-\frac{1}{b^n}$.

These results may be conveniently arrived at by reducing one angular
point as $A$ of the quadrilateral to rest. The resolved part of all
the forces which act on each particle perpendicular to the straight
line joining it to $A$ will then be zero. The case of three
particles may be treated in the same manner." }

Routh's priority claim does not concern Dziobek's formula $(8)$, which he
does not write. But the formulas he numbered
(1), (2), (3), (4) in  \cite{Routh2} express the
$\theta_i$'s of our vectorial proof. Routh could easily prove $(8)$ with
his vectorial methods.

{\bf 4. More equations for mutual distances.} It is clear from Equations $(5)$ that the quantity
$$\Delta_1\|q-q_1\|^2+\cdots+\Delta_n\|q-q_n\|^2$$
does not depend on the point $q$. So if we set
$s_{ij}=\|q_i-q_j\|^2$ and $t_i=\sum_{j\neq i}\Delta_js_{ij}$ we get
immediately
$$t_1=t_2=\cdots=t_n,\eqno(10)$$
i.e. for any $i$, $j$, $1\leq i<j\leq n$,
$$t_i-t_j=(\Delta_j-\Delta_i)s_{ij}+\sum_{k\neq i,j}\Delta_k
(s_{ik}-s_{jk})=0.\eqno(11)$$ Using this expression we can get a simpler
proof of the following lemma proved in  \cite{Albouy3}.

{\bf Lemma 1.} For a $n$ body central configuration in dimension $n-2$, the
inequality
$$\Bigl(\frac{\Delta_i}{m_i}-\frac{\Delta_j}{m_j}\Bigr)(\Delta_i-\Delta_j)\geq
0$$ holds for any $i$ and $j$, $1\leq i<j\leq n$.

{\sl Proof.} As we assume $a<0$, $s_{ik}-s_{jk}$ has
the sign of $s_{jk}^a-s_{ik}^a$ which is the sign of
$\mu\Delta_k(\Delta_j/m_j-\Delta_i/m_i)$ by $(8)$. The term with ``$\sum$" in $(11)$
has the sign of $\mu(\Delta_j/m_j-\Delta_i/m_i)$. We deduce that
$\Delta_j-\Delta_i$ has the sign of
$\mu(\Delta_i/m_i-\Delta_j/m_j)$. As there must be a pair of
$\Delta_i$'s having opposite signs, we deduce that $\mu<0$. The required
inequality then follows immediately. QED

We can restrict the study to {\sl
normalised central configurations}, choosing the normalisation
$\lambda=M$. As
$a\neq0$, Equation $(4)$ shows that one can obtain a normalised
central configuration from any central configuration by changing its
scale. Normalised central configurations satisfy Equations $(10)$
and, for some $\mu\in\real$,
$$s_{ij}^a-1=\mu\frac{\Delta_i\Delta_j}{m_im_j}.\eqno(12)$$
We said that $(\Delta_1,\dots,\Delta_n)$ is only defined up to a
factor. We can fix this factor and remove the parameter $\mu$, which is negative
according to the previous proof. To an
$(n-2)$-dimensional normalised central configuration we associate the
unique
$(\Delta_1,\dots,\Delta_n)\in\real^n$ satisfying $(5)$,
$\Delta_i\Delta_j<m_im_j$ and
$$s_{ij}=\Bigl(1-\frac{\Delta_i\Delta_j}{m_im_j}\Bigr)^\alpha,\quad \hbox{where  }
\alpha=1/a.\eqno(13)$$ These
coordinates in turn determine the central configuration up to an
isometry: from the $\Delta_i$'s we find the mutual distances through
Expression $(13)$.

{\bf 5. The main lemmas}

{\bf Lemma 2.} Set $\rho_1<\rho_2\leq 0$, $\rho_1\rho_2<1$,
$\rho\geq 0$. Choose any $\alpha<0$, set
$s_{12}=(1-\rho_1\rho_2)^\alpha$, $s_{13}=(1-\rho_1\rho)^\alpha$,
$s_{23}=(1-\rho_2\rho)^\alpha$ and
$A(\rho_1,\rho_2,\rho)=(\rho_2-\rho_1)s_{12}-(\rho_1+\rho_2)(s_{13}-s_{23})$.
Then $A(\rho_1,\rho_2,\rho)>0$.

{\sl Proof.} We first minimise $A$ with respect to $\rho$. It is enough to minimise
$g(\rho)=s_{13}-s_{23}=(1-\rho_1\rho)^\alpha-(1-\rho_2\rho)^\alpha$. We write $g'(\rho)=0$,
i.e.\ $\rho_1(1-\rho_1\rho)^{\alpha-1}=\rho_2(1-\rho_2\rho)^{\alpha-1}$. Taking the power $1/(\alpha-1)$,
the equation becomes linear in $\rho$. There is at most one root, given by
$$\rho_0=\frac{(-\rho_1)^{1/(\alpha-1)}-(-\rho_2)^{1/{(\alpha-1)}}}{\rho_1(-\rho_1)^{1/(\alpha-1)}-
\rho_2(-\rho_2)^{1/{(\alpha-1)}}}.$$ To simplify this expression we
set $u=1/(\alpha-1)$, i.e.\ $u+1$=$\alpha/(\alpha-1)$. Since
$\alpha<0$, we have $-1<u<0$. We suppose also that
$q=\rho_2/\rho_1$. It is easy to see that $0<q<1$. So
$$\rho_1\rho_0=\frac {1-q^u}{1-q^{u+1}}<0.$$ By plugging $\rho_0$
into $g(\rho)$ and using the definition of $q$, we obtain the
minimal value of $s_{13}-s_{23}$:
$$g(\rho_0)=\Bigl(\frac{q^u-q^{u+1}}{1-q^{u+1}}\Bigr)^\alpha-\Bigl(\frac{1-q}{1-q^{u+1}}\Bigr)^\alpha
=(q^{1+u}-1)\Bigl(\frac{1-q}{1-q^{u+1}}\Bigr)^\alpha.$$ In the
introduced new variables, the problem of minimising $A$ can be
converted into that of minimising
$$B=\frac {A}{(-\rho_1)}=(1-q)(1-\rho_1^2q)^\alpha+(1+q)g(\rho).$$
The variables are $q$, $\rho_1$ and $\rho$. Taking minimiser of $B$
in $\rho$ is the same as replacing $g(\rho)$ by $g(\rho_0)$ above.
Note that the second term in this expression of $B$ does not depend
on $\rho_1$. So, to minimise $B$ in $\rho_1$ it suffices to set
$\rho_1=0$, as one can see immediately. Then we have
$$B> C=(1-q)+(1+q)g(\rho_0)=(1-q)\Bigl(1-\frac{(1+q)(1-q)^{\alpha-1}}{(1-q^{1+u})^{\alpha-1}}\Bigr).$$
To prove $C>0$, we only need to prove that
$(1-q^{1+u})^{1-\alpha}(1+q)(1-q)^{\alpha-1}<1$. By raising
to the power $u=1/(\alpha-1)$, the formula changes to
$$\frac{(1+q)^u(1-q)}{1-q^{1+u}}>1.$$
So it suffices to conclude that this inequality holds for $q\in]0,1[$
and $u\in ]-1,0[$. We will prove
$$f(q)=(1+q)^u(1-q)+q^{1+u}-1>0.$$ One writes $1-q=-(1+q)+2$ and
calculates the derivative
$f'(q)=-(1+u)(1+q)^u+2u(1+q)^{u-1}+(1+u)q^u$. By factoring out
$(1+q)^u$, one gets the factor $-(1+u)+2u(1-x)+(1+u)x^u$. Here we
used the new variable $x=q/(1+q)\in ]0,1/2[$, which satisfies
$(1-x)(1+q)=1$. This factor is a Laguerre trinomial. It has at most
two positive roots. This implies the same for $f'(q)$ and excludes
the possibility of a root of $f$ in $]0,1[$. In fact, since
$f(0^+)=0^+$, $f(1)=0$ and $f'(1)=-2^u+1+u<0$, $f$ having another
root between $0$ and $1$ would imply $f'$ having at least three
positive roots. QED

{\bf Lemma 3.} Substituting the $s_{ik}$ using $(13)$, we
consider $t_i=\sum_{k\neq i}\Delta_k s_{ik}$ as a function of
$\Delta_1,\dots,\Delta_n$ satisfying
$\Delta_1+\dots+\Delta_n=0$ and of the parameters $a<0$, $m_1>0,\dots, m_n>0$.
Let
$\Delta_1/m_1<\Delta_2/m_2\leq 0$,
$\Delta_1\Delta_2<m_1m_2$, $\Delta_i\geq 0$  for $i\geq 3$. Then
$m_1\geq m_2\Longrightarrow t_1>t_2$.

{\sl Proof.} We consider $(11)$ and renumber the particles in such a
way that $s_{13}-s_{23}\leq s_{14}-s_{24}\leq\cdots\leq
s_{1n}-s_{2n}$. We get
$t_1-t_2\geq Z$ where $Z=(\Delta_2-\Delta_1)s_{12}+(\Delta_3+\cdots+\Delta_n)(s_{13}-s_{23})
=(\Delta_2-\Delta_1)s_{12}-(\Delta_1+\Delta_2)(s_{13}-s_{23})$.
We set $s_1=-s_{12}-s_{13}+s_{23}$ and $s_2=s_{12}-s_{13}+s_{23}$.
By $(13)$  $s_{12}>1$ and $s_{23}<1$.
Then $s_1<0$ and $Z/m_2=(\Delta_1/m_2)s_1+(\Delta_2/m_2)s_2\geq
(\Delta_1/m_1)s_1+(\Delta_2/m_2)s_2=A(\Delta_1/m_1,\Delta_2/m_2,\Delta_3/m_3)>0$
by Lemma 2. QED

{\bf 6. Proof of Theorem 1.} Consider the first claim. The ``only
if" part is easy and does not use the convexity of the
configuration. Under the symmetry hypothesis $s_{13}=s_{23}$ and
$s_{14}=s_{24}$. Using $(13)$ this gives
$\Delta_1/m_1=\Delta_2/m_2$. But the symmetry also implies
$\Delta_1=\Delta_2$ so $m_1=m_2$. We pass to the ``if" part. We
assume $m_1=m_2$. The convexity with particles 1 and 2 on a diagonal
means, without loss of generality, $\Delta_1\leq 0$, $\Delta_2\leq
0$, $\Delta_3\geq 0$ and $\Delta_4\geq 0$. Assuming
$\Delta_1<\Delta_2$, Lemma 3 applies and gives $t_1>t_2$. This
contradicts $(10)$: there is no planar central configuration with
these $\Delta_i$'s. The inequality $\Delta_2<\Delta_1$ is excluded
in the same way. So $\Delta_1=\Delta_2$ and the configuration is
symmetric according to $(8)$ or $(13)$.

Concerning the other claims, what remains to prove reduces to
$m_1<m_2\Longleftrightarrow
\Delta_1<\Delta_2<0$. This is obtained from Lemma 3, which gives
$\Delta_1<\Delta_2<0\Longrightarrow m_1<m_2$ and
$\Delta_2<\Delta_1<0\Longrightarrow m_2<m_1$. Note that here we have
used the first part of the theorem and Lemma 1. QED

{\bf 7. Higher dimensional results}

\nobreak
{\bf Theorem 2.} Let five particles $(q_1,q_2,q_3,q_4,q_5)$  form a
central configuration, which is a convex three-dimensional
configuration having $[q_1,q_2]$ as the diagonal. This configuration
is symmetric with respect to the plane $[q_3,q_4,q_5]$ if and only
if $m_1=m_2$. Also $m_1<m_2$ if and only if
$|\Delta_{1345}|<|\Delta_{2345}|$, where $\Delta_{ijkl}$ is the
oriented volume of the tetrahedron $[q_i,q_j,q_k,q_l]$.

\begin{figure*}[h!] %5
\centerline{\includegraphics [width=110mm] {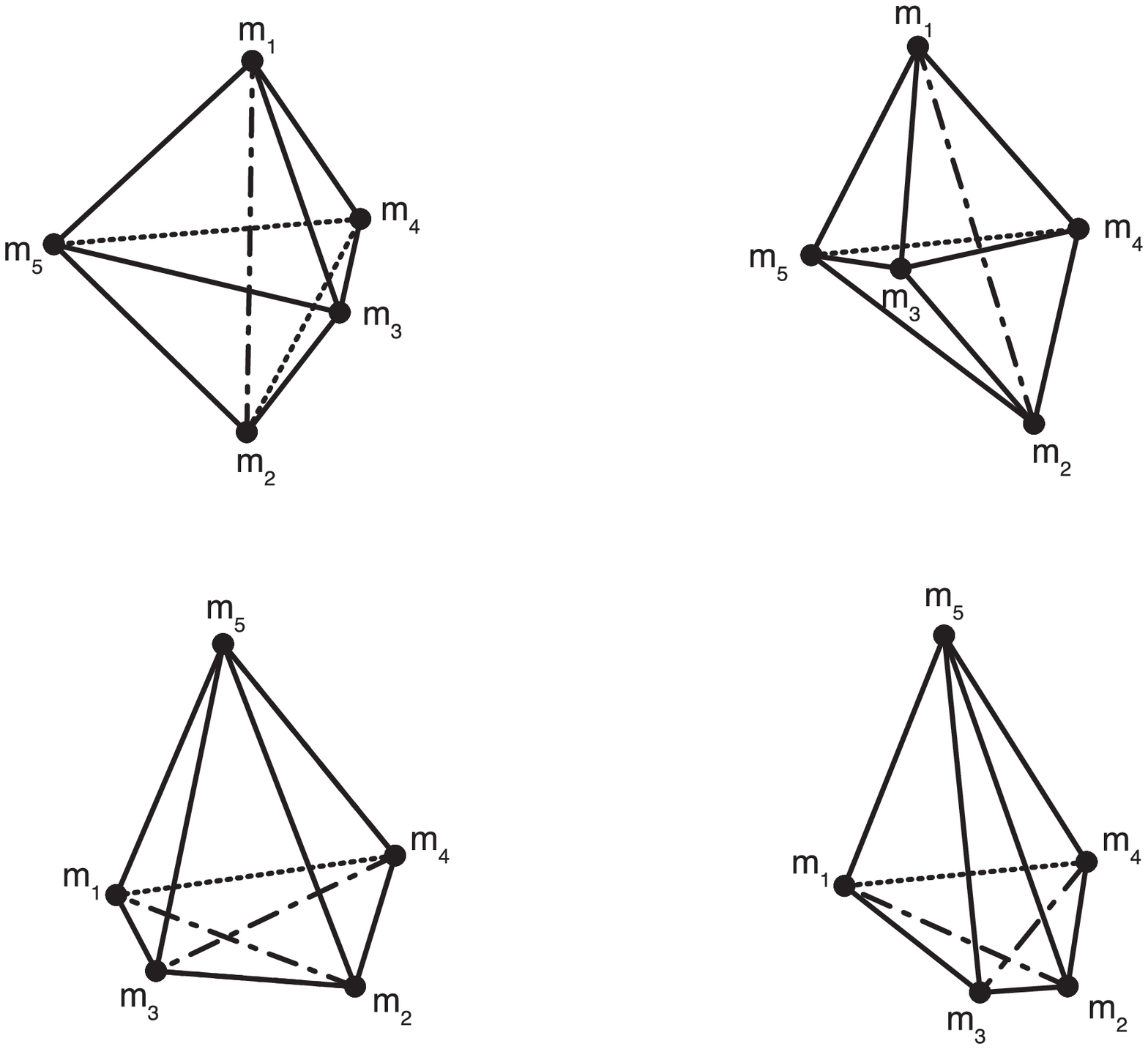}}
\caption{Spatial 5-body convex configurations}
\centerline{Left: with a plane of symmetry. Right: asymmetric.}
\end{figure*}

We say ``the diagonal" because a generic convex 3D configuration of
five points has a unique diagonal. But if four particles are
coplanar there are two diagonals in this plane (see
Fig.\ 2). The above theorem is valid for this kind of diagonals also.
The proof is similar to that of Theorem 1.

We obtained the symmetry but do not know the existence and the uniqueness of the central configuration. We do not know if
the number of symmetric solutions is finite. About the existence, the nice
arguments in Xia's paper \cite{Xia}  show that there exists a convex
central configuration, but they don't show there exits one having
$[q_1,q_2]$ as the diagonal. In contrast with four body case, there
are paths in the space of convex configurations going from
configurations having $[q_1,q_2]$ as the diagonal to, let us say,
configurations having $[q_3,q_4]$ as the diagonal. The limiting
configuration is such that $(q_1,q_2,q_3,q_4)$ form a convex planar quadrilateral
configuration. This shape is possible for a central configuration.

In term of the $\Delta_i$'s, such a path starts with, for example, a hexahedron with $\Delta_1<\Delta_2<0<\Delta_5<\Delta_3<\Delta_4$, passes through a pyramidal configuration where $\Delta_5=0$ and finishes with a hexahedron with $\Delta_1<\Delta_2<\Delta_5<0<\Delta_3<\Delta_4$.

The higher dimensional version, with $n$ particles in dimension
$n-2$, is also true, if the convex configuration is obtained by
gluing two simplices by a common hyperface. We have to be careful
that from $n=6$ this is not the only way to get a convex
configuration. We can think of this in term of the signs of the
$\Delta_i$'s. If only one of them is negative, we are in a
non-convex case. If exactly two of them are negative, we are in the
convex case where our Lemma 3 may apply. If there are exactly
three negative $\Delta_i$'s, the configuration is convex but
our Lemma does not apply. Another way to get some geometrical
intuition is to think of the configuration as a simplex plus a
point, this point being added nearby but outside a hyperface, or
nearby but outside a lower dimensional face.

\end{document}